\documentclass[showpacs,showkeys,amsmath,amssymb,preprint,nofootinbib]{revtex4}
\usepackage{graphicx}
\usepackage{epsfig}
\usepackage{bm}
\usepackage{float}
\usepackage[T1]{fontenc}
\usepackage[utf8]{inputenc}
\usepackage{graphicx}
\usepackage{bm}
\usepackage{amsmath}
\usepackage{xcolor}
\DeclareUnicodeCharacter{2212}{\textendash}
\def\beq{\begin{eqnarray}}

\def\eeq{\end{eqnarray}}

\begin{document}

\title[RMC]{Matter production effects and interacting scenario within a reconstructed mimetic cosmology for late times}

\author{V\'ictor H. C\'ardenas$^{a,}$}
\email{victor.cardenas@uv.cl}

\author{Miguel Cruz$^{b,}$}
\email{miguelcruz02@uv.mx}

\author{Samuel Lepe$^{c,}$}
\email{samuel.lepe@pucv.cl}

\affiliation{$^a$Instituto de F\'{\i}sica y Astronom\'ia, Universidad de Valpara\'iso, Gran Breta\~na 1111, Valpara\'iso, Chile\\
$^b$Facultad de F\'\i sica, Universidad Veracruzana 91000, Xalapa, Veracruz, M\'exico\\
$^c$Instituto de F\'\i sica, Facultad de Ciencias, Pontificia Universidad Cat\'olica de Valpara\'\i so, Av. Brasil 2950, Valpara\'\i so, Chile}

\date{\today}

\begin{abstract}
In this work we explore two possible scenarios that can be considered to extend a recent proposed model by the authors known as {\it reconstructed mimetic cosmology}. This study is complemented with an statistical analysis for each case. The first scenario considers the inclusion of matter production as a possible source of cosmic expansion in the reconstructed mimetic model, at effective level was found that this construction can cross the phantom divide, the model evolves from quintessence to phantom dark energy. The second scenario corresponds to a construction of an interacting scheme for the dark sector which is described by the unified mimetic model. The resulting interaction term (not imposed by an Ansatz), $Q$, exhibits changes of sign leading to the violation of the second law along the cosmic evolution and non adiabaticity; the temperatures for the components of the dark sector are computed and such components are shown to be out of thermal equilibrium.
\end{abstract}



\maketitle

\section{Introduction}
\label{sec:intro}
General Relativity is the most successful theoretical framework describing the gravitational interaction since its foundations have been corroborated by several experiments. However, in order to describe the current state of the Universe we require a scheme beyond General Relativity: the addition of two extra components must be considered. These elements are known as dark matter and dark energy, each of them with a specific task; the clustering of matter and the driving of the accelerated cosmic expansion, respectively. The origin and nature of the dark sector are still surrounded by several questions, then both subjects become part of the fundamental problems to be solved by the contemporary physics. In this sense, modifications and/or extensions of General Relativity provide scenarios where the aforementioned components could have a natural explanation.\\

The mimetic gravity description \cite{mukhanov1} provides a geometric origin for cold dark matter by considering that a conformal degree of freedom of gravity is encoded in an extra scalar field. Then, the resulting model extends the symmetries of General Relativity since the invariance under conformal transformations is added but also has the interesting feature of maintaining second order equations of motion. Few time after the introduction of mimetic gravity it was found that a description for the dark sector could be provided by a minimal extension of the mimetic approach, i.e., inclusion of a potential function for the conformal scalar field and the consideration of a Lagrange multiplier at action level \cite{mukhanov2}. Some other extensions for mimetic gravity can be found in the literature, for instance, in Ref. \cite{mg1} a vector field is considered instead a scalar field, this modification reformulates the mimetic approach as a theory free from certain instabilities and pathologies inherent to the scalar field. See also Ref. \cite{exor}, where an {\it exorcised} version of mimetic gravity is proposed by analyzing the perturbations of the theory. An interesting review on the role of mimetic gravity and some of its modifications in cosmology for late times can be found in \cite{mg2}, see also \cite{saridakis} for a dynamical approach. In fact, the mimetic scenario is also viable to describe the early universe, the role of the mimetic scalar field and the curvature invariant in the context of loop quantum gravity is discussed in Refs. \cite{mg3, mg4}. The mimetic approach was also adapted for higher dimensional objects also known as branes \cite{mg5}.\\

Another interesting aspect of mimetic gravity is the avoidance of singularities during the cosmic evolution. This could be achieved by considering some conditions on the curvature scalar. In Ref. \cite{mg6} is exposed the case where Universe under the mimetic approach exhibits an oscillatory (bouncing) nature, being the spatial curvature the responsible for this behavior. This result shows that in fact mimetic gravity has significant deviations from General Relativity, where the collapse phase ends inevitably in a curvature singularity. For models of modified gravity in the mimetic approach see \cite{Nojiri:2014zqa, Odintsov:2015cwa, Astashenok:2015haa, Matsumoto:2015wja, Nojiri:2016vhu, Odintsov:2016oyz, Nojiri:2016ppu, Odintsov:2015wwp}. Finally, mimetic gravity could be a viable alternative to shed some light on the dark sector, it provides a tentative theoretical laboratory to prove some possible interactions between dark matter and other particles of the standard model such as baryons and photons. In Ref. \cite{interact} was found that these kind of interactions can be possible only with derivative couplings of the mimetic field, but, these interactions can have a direct impact on observables of our Universe such as the CMB.\\  

The structure of the work is the following: Section \ref{sec:recons} is devoted to provide a brief description of the reconstructed scenario proposed by the authors for mimetic gravity with the consideration of the Chevallier-Polarski-Linder (CPL) parametrization for the parameter state. In Section \ref{sec:mp} the matter production approach with a specific production rate is considered within the reconstructed mimetic cosmology. As we will see below, after performing the statistical analysis it is possible to show that the model admits a quintessence dark energy at present time but eventually the model evolves to a over accelerated stage, i.e., phantom scenario. We end this section discussing the case of cold dark matter production. In Section \ref{sec:interaction} we discuss some cosmological implications for the interaction scenario emerging from the mimetic cosmology. We implement a known method to construct this interaction scheme. Based on the data analysis we show that the resulting interaction term exhibits changes of sign and from some thermodynamics considerations we establish the region of validity for this interacting scenario, resulting valid only from past until present time, therefore the future singularity induced by the CPL parametrization is avoided. The cosmic expansion for this model is not adiabatic and the the second law at present time is guaranteed only under certain conditions. This interacting scenario is out of thermal equilibrium. In Section \ref{sec:final} we give the final comments of our work. In this work we will consider $8\pi G=c=k_{B}=1$ units.   
 
\section{Reconstructed mimetic cosmology}
\label{sec:recons}
In this section we give some highlights of the reconstructed scenario for mimetic cosmology discussed by the authors in Ref. \cite{rmc}. The action describing the mimetic field, $\phi$, involves a Lagrange multiplier as follows
\begin{equation}
    S = \int_{M} d^{4}x\sqrt{-g}\left\lbrace \frac{R(g)}{2}-\frac{\lambda}{2}\left[g^{\mu \nu}\partial_{\mu}\phi\partial_{\nu} \phi -1\right] +\mathcal{L}_{m} - V(\phi) \right\rbrace,
    \label{eq:modified}
\end{equation}
we also consider the inclusion of a potential for the mimetic field namely, $V(\phi)$ and $\mathcal{L}_{m}$ represents the Lagrangian density for matter fields. The consideration of a Lagrange multiplier enforces an important constraint that must satisfied by the mimetic field, $g^{\mu \nu }\partial _{\mu }\phi \partial _{\nu }\phi =-1$, as stated in Ref. \cite{mukhanov1}. An interesting generalization of model (\ref{eq:modified}) is given by replacing the scalar curvature, $R$, of the gravitational sector by some appropriate function of such scalar, i.e., $f(R)$ gravity; in fact, this latter scenario can be enriched with the consideration of higher-order curvature terms. In Ref. \cite{epjcref} the gravity sector of (\ref{eq:modified}) is described by a function of the form, $f(R,R_{\mu \nu}R^{\mu \nu})$, for a specific function, $f$; this scheme allows to reconstruct the Lagrange multiplier and the mimetic field potential, in other words, an Ansatz-free description. Another relevant feature of this higher-curvature mimetic description is its capability to unify various stages of cosmic evolution, the transition from matter domination to an accelerated stage and also bounce cosmology can be obtained under certain considerations. This kind of unification was also discussed in the context of a power law $f(R)$ mimetic gravity in Ref. \cite{Nojiri:2016vhu}, where was also found that according with the choice of the potential and Lagrange multiplier the model leads to nearly quintessential or nearly phantom late time acceleration. See also Ref. \cite{review} for other possible modifications of the gravity sector in the mimetic description. The variation of action (\ref{eq:modified}) with respect to the metric provides the following equations of motion,
\begin{equation}
G_{\mu \nu} = T_{\mu \nu}+\lambda \partial_{\mu}\phi \partial_{\nu} \phi - g_{\mu \nu}\left[\frac{\lambda}{2}(\partial^{\alpha}\phi \partial_{\alpha} \phi -1) + V(\phi)\right],
\label{eq:einst}
\end{equation}%
where $G_{\mu \nu }$ and $T_{\mu \nu }$ are the Einstein tensor and the matter energy-momentum tensor, respectively. On the other hand, the variation of the action with respect to the mimetic field leads to a modified version of the Klein-Gordon equation
\begin{equation}
    \nabla^{\mu}(\lambda \partial_{\mu}\phi) - \frac{dV(\phi)}{d \phi} = 0,
    \label{eq:gordon}
\end{equation}
note that the constraint mentioned above for the mimetic field can be obtained by performing a variation of the action with respect to the Lagrange multiplier. If we consider a FLRW configuration, the homogeneity and isotropy of spacetime is preserved by taking the scalar field as, $\phi = \phi(t)$. Then, according to the constraint $g^{\mu \nu }\partial _{\mu }\phi \partial _{\nu }\phi =-1$ we obtain, $\dot{\phi}^{2} = 1$, where the dot stands for derivatives with respect to cosmic time. The previous expression leads immediately to $\phi(t) = t$. For this geometry, the Einstein equations take the following form
\begin{align}
    & 3H^{2} = \rho + \lambda + V, \label{eq:friedmann}\\
    & 2\dot{H} + 3H^{2} = V - p,
    \label{eq:accel}
\end{align}
where $H$ is the Hubble parameter defined as $H:=\dot{a}/a$, being $a$ the scale factor. From these last results it is possible to identify the potential for the mimetic field as a pressure, $p_{\mathrm{mg}} = - V$. Then, from the Friedmann constraint we can also write the energy density of the mimetic gravity as, $\rho_{\mathrm{mg}} = \lambda + V$. If we consider this relation between the energy density and the Lagrange multiplier, we can integrate the expression (\ref{eq:gordon}), yielding \cite{rinaldi}
\begin{eqnarray}
\rho_{\mathrm{mg}}(a) &=& \frac{c}{a^{3}} + \frac{3}{a^{3}}\int^{t}_{t_{0}}a^{3}(t')H(t')V(t')dt' \label{eq:densitya}\\ &=& \frac{c}{a^{3}} + \frac{3}{a^{3}}\int a^{2}V da,
\label{eq:densityb}
\end{eqnarray}
therefore from this latter result it is clear the role that the potential $V$ plays in this description; it is the responsible of producing a deviation from the standard result $a^{-3}$, for the matter sector and also for the emergence of a dark energy behavior in the mimetic approach. For a vanishing potential the results obtained in \cite{mukhanov1} are recovered. In general, these deviations from the standard behavior for the matter sector can be attributable to a possible interaction of this sector with some other component. We will return to this point later. Then, in Ref. \cite{rmc} was found that by means of a CPL parametrization for the parameter state 
\begin{equation}
    \omega_{\mathrm{mg}}(z) = \omega_{0}+\omega_{a}\frac{z}{1+z},
    \label{eq:cpl}
\end{equation}
one gets for the energy density
\begin{equation}
    \rho_{\mathrm{mg}}(z) = \rho_{\mathrm{mg,0}}(1+z)^{3(1+\omega_{0}+\omega_{a})}\exp\left(-3\omega_{a}\frac{z}{1+z} \right).
    \label{eq:density}
\end{equation}
Then, using the equation (\ref{eq:densityb}) and the standard relation between the redshift and the scale factor, $1+z = a_{0}/a$, it is possible to find an explicit expression for the potential scalar field that leads to energy density as given in (\ref{eq:density}), one gets
\begin{equation}
    V(z) = - \rho_{\mathrm{mg,0}}(1+z)^{3(1+\omega_{0}+\omega_{a})}\left[\omega_{0}+\omega_{a}\frac{z}{1+z} \right]\exp \left(-3\omega_{a}\frac{z}{1+z} \right),
    \label{eq:potential}
\end{equation}
from now on we choose the value $a_{0} = 1$. Notice that $\omega_{\mathrm{mg}}(z=0) = \omega_{\mathrm{mg}}(a=a_{0}) = \omega_{0}$, which will represent the value at present time for the parameter state (\ref{eq:cpl}). Finally, from the expression, $\rho_{\mathrm{mg}} = \lambda + V$, we find for the Lagrange multiplier
\begin{equation}
    \lambda(z) = \rho_{\mathrm{mg,0}}(1+z)^{3(1+\omega_{0}+\omega_{a})}\left[1+\omega_{0}+\omega_{a}\frac{z}{1+z} \right]\exp\left(-3\omega_{a}\frac{z}{1+z} \right),
    \label{eq:multiplier}
\end{equation}
and this expression represents the {\it deviated} matter sector in this construction. An important feature of this scenario is that no Ansatze for $V$ are needed since the CPL parametrization allowed its construction.   

\section{First scenario: matter production}
\label{sec:mp}
In order to visualize the performance of the reconstructed mimetic gravity in an enriched scenario, we consider the gravitational production of matter. In this context the matter production is due to the expansion of the Universe \cite{prod1}. Besides, for this scenario the production of particles is a consequence only of the dynamical gravitational background, i.e., the created particles interact only with gravity, therefore their abundance is determined by the mass of the particles uniquely. This theoretical scenario has been the subject of various tests to explain the nature of dark matter, despite the fact that dark matter only interacts with gravity, there could be various mechanisms for its production \cite{prod2}. For a matter production scheme we have the following conservation equations for the matter sector in a FLRW spacetime
\begin{equation}
\dot{n} + 3H n = n \Gamma, \hspace{1cm} \dot{\rho} + 3H (\rho + P) = 0,
\label{eq:conseqs}
\end{equation}
where $\Gamma >0$, $\Gamma < 0$ acts like a source or sink of particles, respectively; $n$ is the particle number density and $P = p + p_{c}$. Here $p_{c}$ accounts for the pressure from matter creation and is defined as 
\begin{equation}
    p_{c} = - \frac{\rho+p}{3H} \Gamma,
    \label{eq:pc}
\end{equation}
where we have considered an adiabatic expansion for the Universe, $\dot{S} = 0$, being $S$ the entropy. In general, from the Gibbs law and Eqs. (\ref{eq:conseqs}) we can write $nT\dot{S}=-3 H p_{c} - (\rho + p)\Gamma$, where $T$ represents the temperature of the fluid. In this case the Friedmann constraint reads, 
\begin{equation}
    3H^{2} = \rho_{\mathrm{m}} + \rho_{\mathrm{mg}}
    \label{eq:comb1}
\end{equation}
by the subscript $\mathrm{m}$ we denote the produced (created) matter. Then, the acceleration equation for this scenario results
\begin{eqnarray}
    \dot{H} + H^{2} = -\frac{1}{6}(\rho + 3p) &=& -\frac{1}{6}\left[\rho_{\mathrm{m}} + \rho_{\mathrm{mg}}+3(p_{c} + p_{\mathrm{mg}}) \right] \nonumber \\ &-&\frac{1}{6}\left[ \rho_{\mathrm{m}} + \rho_{\mathrm{mg}} + 3 \left\lbrace -\frac{ \rho_{\mathrm{m}}\Gamma}{3H}+ \omega_{\mathrm{mg}}\rho_{\mathrm{mg}} \right\rbrace\right],
    \label{eq:comb2}
\end{eqnarray}
notice that we restrict ourselves to the case of created dark matter, for this case we have $p=0$, and for the pressure coming from the mimetic contribution we considered a barotropic equation of state, $p = \omega \rho$. If we combine the Eqs. (\ref{eq:comb1}) and (\ref{eq:comb2}) we can obtain the following continuity equation (\ref{eq:conseqs}) 
\begin{equation}
    \dot{\rho}+3H(1+\omega_{\mathrm{eff}})\rho = 0,
\end{equation}
where $\rho$ is the total energy density $\rho := \rho_{\mathrm{m}} + \rho_{\mathrm{mg}}$ and 
\begin{eqnarray}
    \omega_{\mathrm{eff}}(z) = - \frac{ \rho_{\mathrm{m}}\Gamma}{3H\rho}+\frac{\omega_{\mathrm{mg}}\rho_{\mathrm{mg}}}{\rho} &=& -\frac{\Gamma}{3H(z)}\left[\frac{1}{1+\rho_{\mathrm{mg}}(z)/\rho_{\mathrm{m}}(z)} \right]+\omega_{\mathrm{mg}}(z)\left[\frac{1}{1+\rho_{\mathrm{m}}(z)/\rho_{\mathrm{mg}}(z)} \right] \nonumber \\ &=& - \frac{\mathcal{G}}{3H}.
    \label{eq:effective}
\end{eqnarray}
where we have defined $\mathcal{G} := \Gamma[1/(1+\rho_{\mathrm{mg}}/\rho_{\mathrm{m}})] + [1/(1+\rho_{\mathrm{m}}/\rho_{\mathrm{mg}})]$. From the previous expression for the effective parameter state we can observe that if, $\rho_{\mathrm{mg}} = 0$, we recover the following equation
\begin{equation}
    \omega_{\mathrm{eff}}(z) = - \frac{\Gamma}{3H}.
    \label{eq:effective1}
\end{equation}
This form for the effective parameter state was also studied in Ref. \cite{us, generalrate, generalrate1}. Then the $\mathcal{G}$-term plays the role of a generalized particle production rate. We will consider as expansion rate the following normalized Hubble parameter, which arises from the consideration of matter production effects and mimetic gravity
\begin{equation}\label{eq:edz2}
    E(z) = \sqrt{\Omega_{\mathrm{m,0}}\left[ \xi+\left( 1-\xi \right) \left( 1+z\right) ^{3\left( 1-\delta
\right) /2}\right] ^{\frac{2}{1-\delta }} + (1-\Omega_{\mathrm{m,0}})(1+z)^{3(1+\omega_{0}+\omega_{a})}\exp\left(-3\omega_{a}\frac{z}{1+z} \right)},
\end{equation}
where $E(z) := H(z)/H_{0}$ and expression (\ref{eq:density}) was considered. Notice that in the previous expression we also considered the normalization condition, $\Omega_{\mathrm{m,0}}+\Omega_{\mathrm{mg,0}}=1$. In order to write the normalized Hubble parameter, for the matter production sector we focused on the following form for the particle production rate
\begin{equation}
\Gamma = 3\xi H_{0}\left(\frac{H}{H_{0}} \right)^{\delta},   
\end{equation}
being $\xi$ and $\delta$ dimensionless constants and $H_{0}$ represents the Hubble parameter evaluated at present time. This model was proposed in Refs. \cite{generalrate, generalrate1} and contains most of the Ansatze found in the literature for $\Gamma$ if we consider some specific values for $\xi$ and $\delta$. By means of Eq. (\ref{eq:conseqs}) we can write for $\delta \neq 1$ \cite{generalrate, generalrate1}
\begin{equation}
    \rho(z) = \rho_{0}\left[ \xi+\left( 1-\xi \right) \left( 1+z\right) ^{3\left( 1-\delta
\right) /2}\right] ^{\frac{2}{1-\delta }},
\label{eq:mcdensity}
\end{equation}
where $\rho_{0}$ is the value of the density at present time.  For this model we have the following expression for the effective parameter state given in (\ref{eq:effective})
\begin{equation}
\omega_{\mathrm{eff}} =  - \xi(E(z))^{\delta - 1} \left[\frac{1}{1+\rho_{\mathrm{mg}}(z)/\rho_{\mathrm{m}}(z)} \right]+\omega_{\mathrm{mg}}(z)\left[\frac{1}{1+\rho_{\mathrm{m}}(z)/\rho_{\mathrm{mg}}(z)} \right]. 
\label{eq:bestomega}
\end{equation}

Using Eq. (\ref{eq:edz2}) we can test it against observations considering six free parameters: $h, \xi, \delta , \Omega_{m,0}, \omega_0$ and $\omega_a$, where $h=H_{0}/100$. Here we use supernova data and $H(z)$ measurements. For type Ia supernova we use the latest supernova sample, the Pantheon sample \cite{scolnic} consisting in 1048 data points. This set gives us the apparent magnitude at maximum brightness, and the covariance and correlations among the data. The data cover the redshift range $0.01 < z <2.3$. We compute the residuals $\mu - \mu_{th}$ and minimize the quantity
\begin{equation}\label{chi2jla}
\chi^2 = (\mu - \mu_{th})^{T} C^{-1}(\mu - \mu_{th}),
\end{equation}
where $\mu_{th} = 5 \log_{10} \left( d_L(z)/10pc\right) $ gives the theoretical distance modulus, $d_L(z)$ is the luminosity distance, $ C $ is the covariance matrix released in \cite{scolnic}. Because the absolute magnitude and the Hubble parameter are degenerated in the computation of the distance modulus, we marginalize over these nuisance parameters using
\begin{equation}
    \chi^2_{sn} = A + \log \frac{D}{2\pi} - \frac{B^2}{D},
\end{equation}
where $A=(\mu - \mu_{th})^T C^{-1}(\mu - \mu_{th})$, $B = (\mu - \mu_{th})^{T} C^{-1} {\bf 1}$ and $D= {\bf 1}^{T} C^{-1} {\bf 1}$, as is explained in \cite{Conley_2010}. The $H(z)$ measurements are obtained from two methods. The first one uses the differential age (DA) method from \cite{Jimenez:2001gg, Simon_2005, Stern_2010} which is based on measurements of
\begin{equation}
    H(z) = \frac{1}{a}\frac{da}{dt}=-\frac{1}{1+z}\frac{dz}{dt}\simeq -\frac{1}{1+z}\frac{\Delta z}{\Delta t},
\end{equation}
which so far consist in 31 measurements compiled from \cite{Wei:2018cov}. The second method is based on measurements of the line of sight BAO data \cite{Blake_2012, Chuang_2013, Font_Ribera_2014, Delubac_2015, Bautista_2017}, which consist in 26 data points extra. In summary, for this observational probe we have $57$ data points which can constrain the Hubble function $E(z)$ for a given model through the $\chi^2$
\begin{equation}
    \chi_{hz}^2 = \sum_{i=1}^{57} \frac{\left\{ H_i -100 h E(z_i) \right\}^2}{\sigma_i^2},
\end{equation}
where $H_i$ are the values of the Hubble function at redshift $z_i$ measured with error $\sigma_i$. The analysis is performed using a
the public code known as \texttt{emcee} \cite{Foreman_Mackey_2013}. This is a stable, well tested Python implementation of the affine-invariant ensemble sampler for Markov chain Monte Carlo (MCMC) proposed by Goodman \& Weare \cite{2010CAMCS...5...65G}. The output from the chains are visualized using GetDist \cite{Lewis:2019xzd}.\\

\begin{figure}[h!]
    \centering
    \includegraphics[width=14cm]{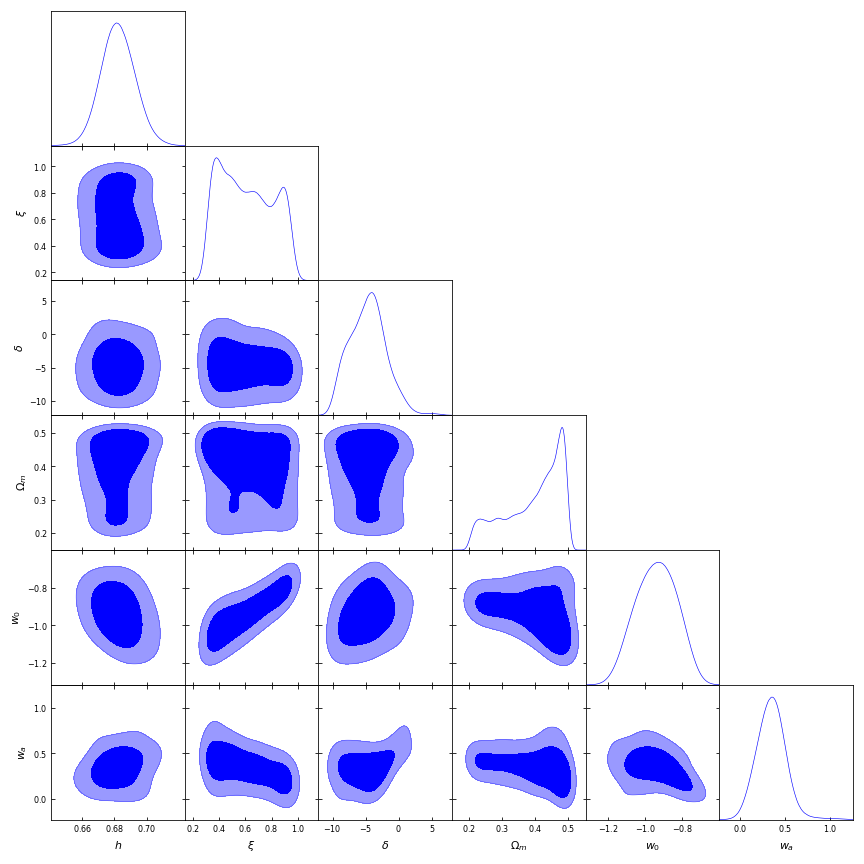}
    \caption{Here we show the results for the posterior 1D and 2D of the model (\ref{eq:edz2}) using both $H(z)$ measurements and supernovae data. }
    \label{fig: fita}
\end{figure}

\begin{figure}[htbp!]
\centering
\includegraphics[width=8cm,height=6cm]{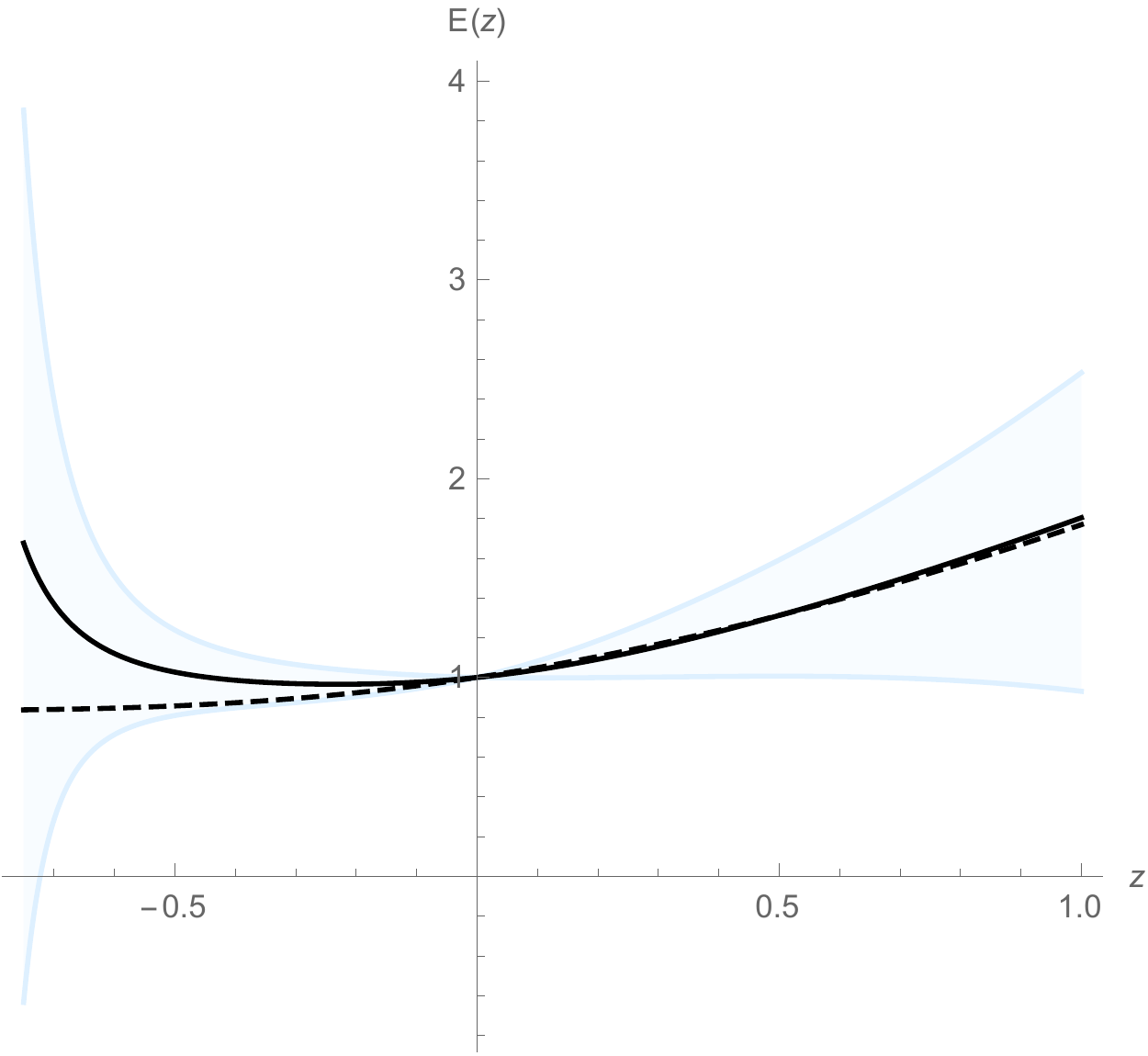}
\includegraphics[width=8cm,height=6cm]{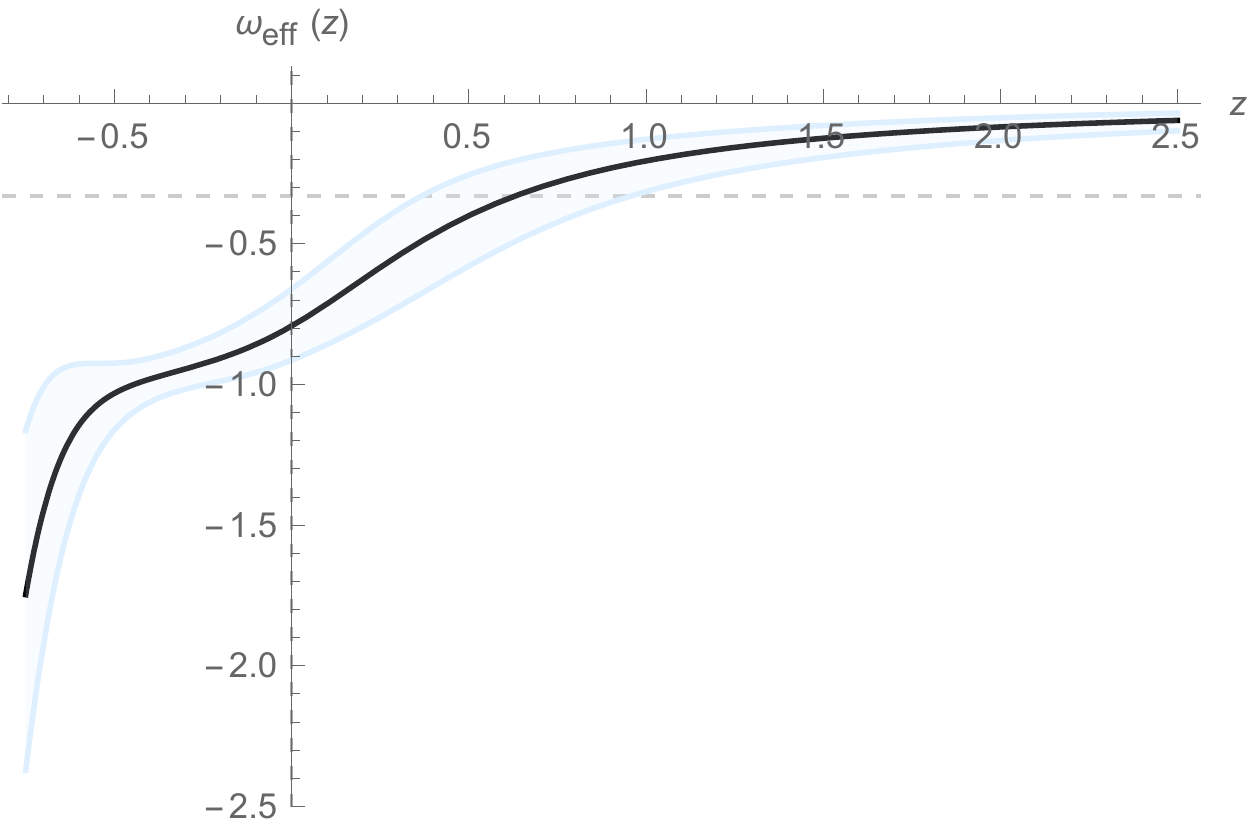}
\caption{The normalized Hubble parameter (left panel) and the effective parameter state (right panel). The black solid line in each case corresponds to the best fit values for the parameters of the model and the shaded regions correspond to the errors propagation. The dashed on the left panel represents the normalized Hubble parameter for the $\Lambda$CDM model.}
\label{fig:MGandCM}
\end{figure}

Assuming all the six parameters free, and without using external priors, we get the results shown in Table (\ref{table: on}).
\begin{table}[b]
\begin{ruledtabular}
\begin{tabular}{lccc}
\textrm{Parameters}&
\textrm{$H(z)$}&
\textrm{SNIa}&
\textrm{$H(z)$+SNIa}\\
\colrule
 $h$ & $0.677 \substack{+0.045\\-0.039}$ & - & $0.682 \substack{+0.011 \\-0.010}$ \\ 
 $\xi$ & $0.60 \substack{+0.22\\-0.21}$ & $0.60 \substack{+0.25\\-0.21}$ & $0.59 \substack{+0.26\\-0.21}$ \\ 
 $\delta$ & $-5.0 \substack{+4.8\\-3.5}$ & $-2.0 \substack{+5.5\\-5.2}$ & $-4.6 \substack{+2.5\\-3.2}$\\
 $\Omega_m$ & $0.40 \substack{+0.08\\-0.12}$ & $0.39 \substack{+0.08\\-0.12}$ & $0.42 \substack{+0.06\\-0.13}$\\
 $\omega_0$ & $-0.92 \substack{+0.17\\-0.22}$ & $-0.91 \substack{+0.09\\-0.11}$ & $-0.94 \substack{+0.12\\-0.12}$\\
 $\omega_a$ & $0.35 \substack{+0.23\\-0.21}$ & $0.55 \substack{+0.43\\-0.50}$ & $0.35 \substack{+0.15\\-0.16}$
\end{tabular}
\end{ruledtabular}
\caption{\label{table: on} Here we display the results of the statistical analysis using $H(z)$ measurements and supernova data. Because the absolute magnitude $\mathcal{M}$ is degenerated with the Hubble parameter $h$, we have marginalized on this parameter for type Ia supernova data.}
\end{table}
The summary of the posterior 1D and 2D probabilities are shown in Fig. (\ref{fig: fita}).
As we see, the Hubble constant $H_0$ take a value close to the Planck value instead the local value obtained from SH0ES. Using both observational probes, the $\delta$ value points close to $-4.6$. Using the best fit values for the parameters of the model and error propagation it is possible to show the behavior of the normalized Hubble parameter (left panel) given in (\ref{eq:edz2}) and the effective parameter state (right panel) written in (\ref{eq:bestomega}). This is displayed in Fig. (\ref{fig:MGandCM}), the horizontal shaded line in the right panel represents the value, $\omega_{\mathrm{eff}} \approx 0.33$. As shown in the figure, the accelerated expansion started at the past and at present time ($z=0$) the fluid behaves as quintessence dark energy. This quintessence behavior is transient, the model evolves to a phantom scenario, $\omega_{\mathrm{eff}} < -1$. It is not shown explicitly in the plot but the effective parameter state diverges at the far future $(z=-1)$. On the other hand, a divergent behavior for $E(z)$ it is also obtained at the far future (not shown explicitly in the plot), therefore a future singularity known as little rip can take place in this description. For the left panel the dashed line represents the normalized Hubble parameter associated to $\Lambda$CDM model, in this case we consider the best fit value for the matter density parameter given as, $\Omega_{\mathrm{m,0}} = 0.3040 \pm 0.0060$, which results from the combination Planck+DES (dark energy survey) \cite{planck}. As can be seen, at the past both functions $E(z)$ almost describe the same evolution, but from present time to future we will have different scenarios.  

\begin{figure}[htbp!]
\centering
\includegraphics[width=8cm,height=6cm]{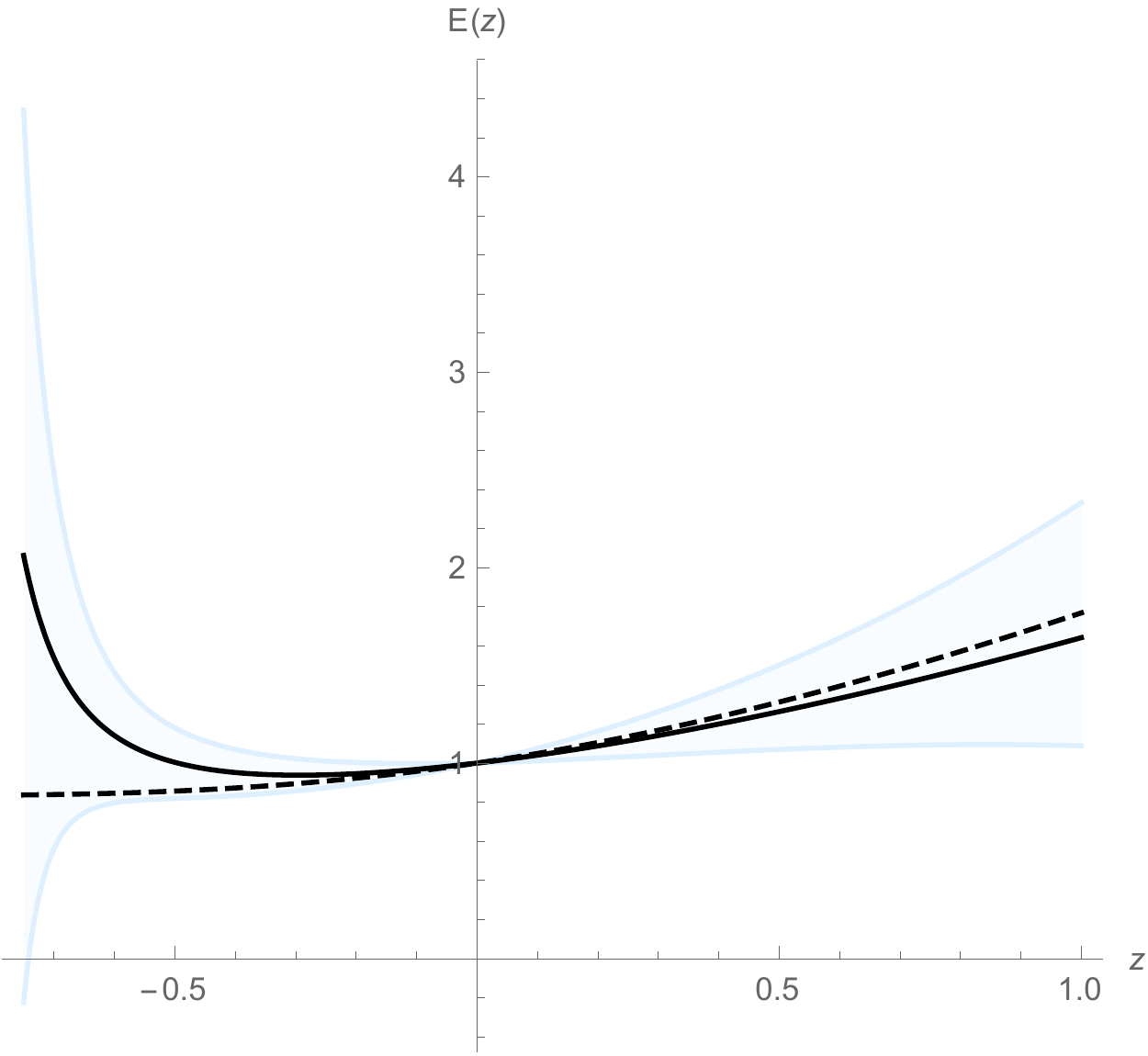}
\includegraphics[width=8cm,height=6cm]{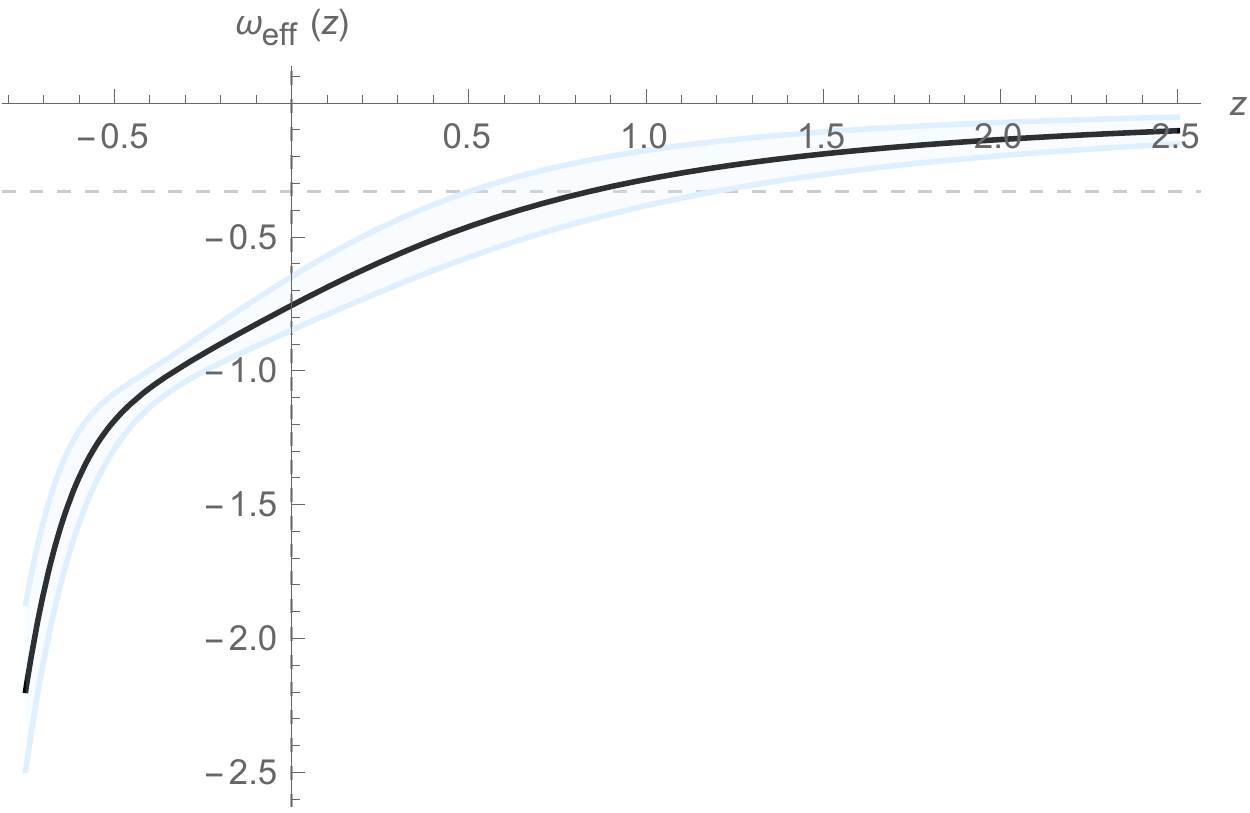}
\caption{On the left panel we can see the comparison between the normalized Hubble parameter of the model and the corresponding one to $\Lambda$CDM model (dashed line) and the effective parameter state (right panel) for the case $\delta = -1$. The black solid line in each panel corresponds to the best fit values for the parameters of the model and the shaded regions correspond to the errors propagation.}
\label{fig:MGandCM2}
\end{figure}

Next, we will focus in the interesting case given by $\delta = -1$, known as creation of cold dark matter that was discussed extensively in Refs. \cite{generalrate, generalrate1, generalrate2}. Then the effective parameter state given in (\ref{eq:bestomega}) takes the following form if we consider the energy density (\ref{eq:density}) obtained for the mimetic gravity sector by means of the CPL parametrization and (\ref{eq:mcdensity}) for the created matter,
\begin{equation}
\omega_{\mathrm{eff}}(z) = \xi(E(z))^{-2} \left[\frac{1}{1+\rho_{\mathrm{mg}}(z)/\rho_{\mathrm{m}}(z)} \right]+\omega_{\mathrm{mg}}(z)\left[\frac{1}{1+\rho_{\mathrm{m}}(z)/\rho_{\mathrm{mg}}(z)} \right].
\label{eq:effcombi3}
\end{equation}
Using the data from both type Ia supernova and $H(z)$ measurements together, assuming $\delta = -1$ we obtain the following best fit for the parameters: $h = 0.694 \substack{+0.010 \\ -0.010}$, $\xi = 0.50\substack{+0.23 \\ -0.15}$, $\Omega_m = 0.39 \substack{+0.08 \\ -0.11}$, $\omega_0 = -0.92 \substack{ + 0.08 \\ -0.09}$, and 
$\omega_a =  0.46 \substack{+ 0.09\\ -0.08}$. On the left panel of Fig. (\ref{fig:MGandCM2}) we show the behavior of the normalized Hubble parameter state (\ref{eq:edz2}) considering $\delta=-1$ and the effective parameter state (\ref{eq:effcombi3}) using the best fit values obtained from the data analysis and error propagation. As in the previous case, the model admits a little rip singularity. From the plot for the effective parameter state can be seen that at present time ($z=0$) we have a quintessence scenario. The region of accelerated cosmic expansion will be eventually driven by a phantom dark energy at effective level as in the previous case. As commented in both cases considered here, the inclusion of matter production leads to quintessence dark energy at present time. This also was obtained in Refs. \cite{us, cm2}, where the matter production was considered. It is worthy to mention that both cases considered could mimic the $\Lambda$CDM model ($\omega_{\mathrm{eff}} = -1$) at some stage of the cosmic evolution.

\section{Second scenario: interaction in the dark sector}
\label{sec:interaction}
As commented in \cite{rmc}, the role of the Lagrange multiplier in the mimetic gravity formulation is to mimic a matter sector contribution that deviates from the standard behavior, $\rho_{\mathrm{m}} \propto (1+z)^{3}$, given that we are considering the inclusion of a potential for the mimetic field. The explicit expression for the Lagrange multiplier is given in (\ref{eq:multiplier}). One could assume that the observed deviation from the usual term can be due to an possible interaction of the matter sector with other components. If we allow that an interaction could exist between the dark matter a dark energy sectors through an interaction term labeled as, $Q$; then for a flat FLRW Universe we can write
\begin{align}
& \rho'_{\mathrm{de}} - 3 \left(\frac{1+\omega_{\mathrm{de}}}{1+z} \right)\rho_{\mathrm{de}} = \frac{Q}{H(z)(1+z)}, \label{eq:q12}\\
& \rho'_{\mathrm{m}} - \left(\frac{3}{1+z} \right)\rho_{\mathrm{m}} = -\frac{Q}{H(z)(1+z)}, \label{eq:q22}
\end{align} 
where the prime stands for derivatives with respect to the cosmological redshift and $H(z)$ is the Hubble parameter; the subscripts $\mathrm{de}$ and $\mathrm{m}$ denote the dark energy and dark matter sectors, respectively. Besides, we have assume for the matter sector $\omega_{\mathrm{m}} = 0$, i.e., cold dark matter. In what follows we will construct the interacting scenario in this description of mimetic gravity by implementing the method given in Ref. \cite{plbgood}. Note that the continuity equations (\ref{eq:q12}) and (\ref{eq:q22}) must be complemented with the Friedmann constraint. If the total energy density is only described by the mimetic gravity given in (\ref{eq:density}), i.e., $\mathcal{L}_{m} =0$ in Eq. (\ref{eq:modified}), and if we also consider that such energy results from the contribution of dark matter and dark energy, we will have
\begin{equation}
    E^{2}(z) = \frac{1}{3H^{2}_{0}}[\rho_{\mathrm{de}}(z) + \rho_{\mathrm{m}}(z)] = \frac{\rho_{\mathrm{mg}}(z)}{3H^{2}_{0}},
    \label{eq:frint}
\end{equation}
therefore, using the Friedmann constraint together with Eqs. (\ref{eq:density}) and (\ref{eq:multiplier}), we can write for the dark energy sector
\begin{eqnarray}
    \Omega_{\mathrm{de}}(z) &=& (1-\Omega_{\mathrm{m,0}})(1+z)^{3(1+\omega_{0}+\omega_{a})}\left\lbrace 1-\frac{\Omega_{\mathrm{m,0}}}{1-\Omega_{\mathrm{m,0}}}\left(\omega_{0}+\omega_{a}\frac{z}{1+z}\right)\right\rbrace \times \nonumber \\ 
    &\times& \exp\left(-3\omega_{a}\frac{z}{1+z} \right), \label{eq:deint}\\
    &=& (1-\Omega_{\mathrm{m,0}})(1+z)^{3(1+\omega_{0}+\omega_{a})}\exp\left(-3\omega_{a}\frac{z}{1+z} \right) + \frac{V(z)}{3H^{2}_{0}},\label{eq:deint3}
\end{eqnarray}
where we have considered the standard definition $\Omega_{\mathrm{i,0}} := \rho_{\mathrm{i,0}}/3H^{2}_{0}$ and the Eq. (\ref{eq:potential}) in the last step. $H_{0}$ is the Hubble constant and both components must obey the normalization condition $\Omega_{\mathrm{de,0}} + \Omega_{\mathrm{m,0}} = 1$.\\

\begin{figure}[htbp!]
\centering
\includegraphics[width=9cm]{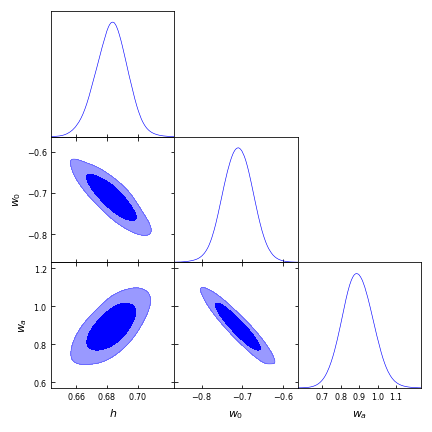}
\caption{Here we show the posteriors of the free parameters using Eq.(\ref{eq:frint}) as the Hubble function and in a joint analysis using type Ia supernova and $H(z)$ measurements. }
\label{fig:mim5all}
\end{figure}

\begin{figure}[htbp!]
\centering
\includegraphics[width=8cm,height=6.5cm]{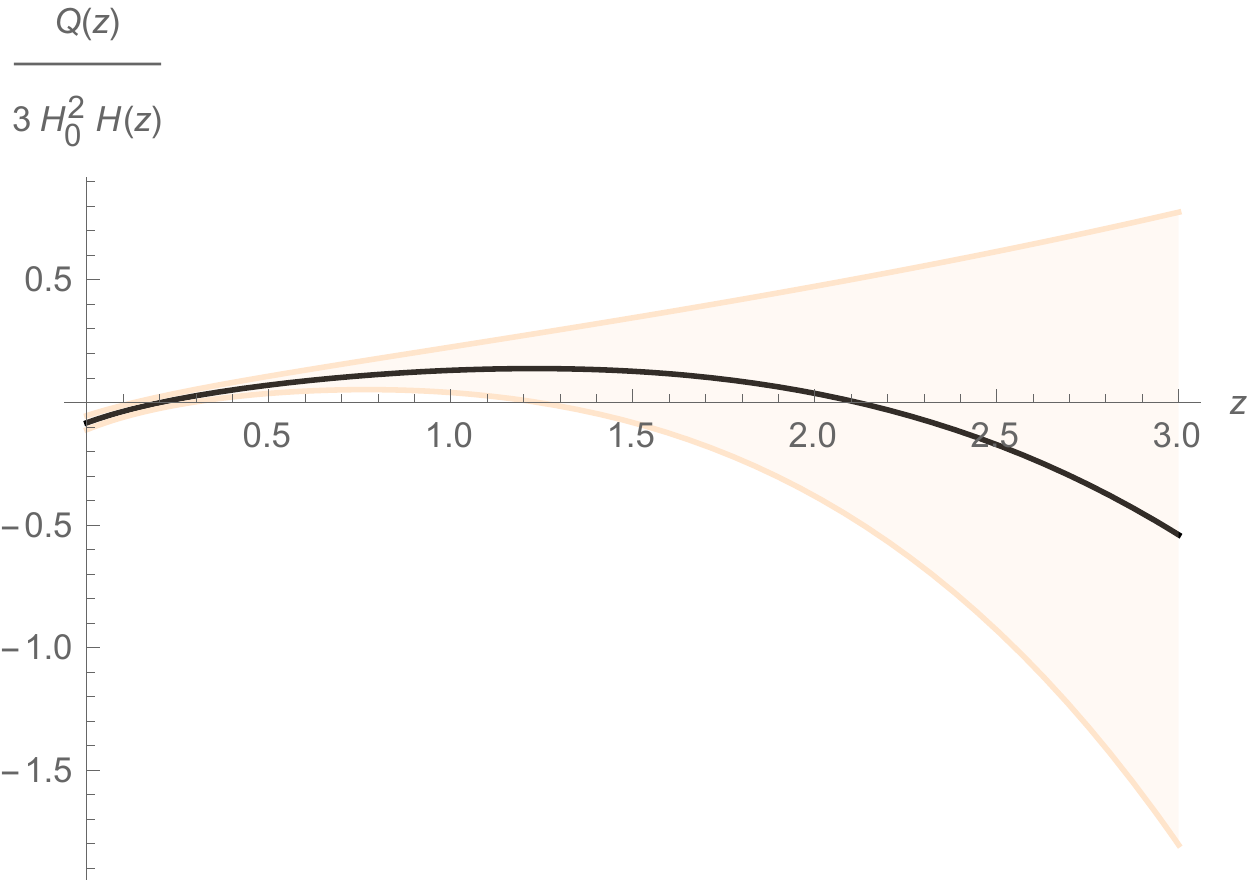}
\caption{$Q$-term. The black solid line corresponds to the best fit values for the parameters of the model and the shaded region corresponds to the errors propagation.}
\label{fig:planck}
\end{figure}

Given (\ref{eq:frint}) it is obvious we have to perform a new test against data, because now the only contribution to the Hubble function is $\rho_{\mathrm{mg}}$ given by (\ref{eq:density}). In this case $h$, $\omega_0$ and $\omega_a$ are the free parameters to fit the data. Using again both type Ia supernova and $H(z)$ measurements we obtain as best fit parameters,  $h = 0.683 \substack{+0.010 \\-0.011 }$, $\omega_0 = -0.711 \substack{+0.036\\-0.037}$, $\omega_a =0.890 \substack{+0.084\\ -0.079}$. The summary of the posteriors are shown in Fig. (\ref{fig:mim5all}).
It is worthy to mention that given the previous results, the mimetic approach can be seen as an unified scenario to describe the dark matter and dark energy content of the universe, the dark matter sector is modeled by the Lagrange multiplier and as can be seen in Eq. (\ref{eq:deint3}), the dark energy sector is related with the potential of the mimetic field. On the other hand, from Eqs. (\ref{eq:q12}), (\ref{eq:q22}) and (\ref{eq:frint}), the parameter state of the dark energy component can be penned as
\begin{equation}
    \omega_{\mathrm{de}}(z) = (1+r(z))\left[-1+\frac{2}{3}(1+z)\frac{E'(z)}{E(z)} \right],
    \label{eq:deint2}
\end{equation}
being $r(z)$ the coincidence parameter defined as the quotient between the energy densities of the dark sector, $r(z) := \Omega_{\mathrm{m}}(z)/\Omega_{\mathrm{de}}(z)$. By computing this quotient between (\ref{eq:multiplier}) and (\ref{eq:deint}) (or Eq. (\ref{eq:deint3})) one gets
\begin{equation}
    r(z) = \frac{\Omega_{\mathrm{m}}(z)}{\Omega_{\mathrm{de}}(z)} = \frac{\Omega_{\mathrm{m,0}}}{1-\Omega_{\mathrm{m,0}}}\left[\frac{1+\omega_{0}+\omega_{a}\frac{z}{1+z}}{1-\frac{\Omega_{\mathrm{m,0}}}{1-\Omega_{\mathrm{m,0}}}\left(\omega_{0}+\omega_{a}\frac{z}{1+z}\right)} \right],
    \label{eq:quotient2}
\end{equation}
note that in this case both densities are related through the function $r(z)$, from the previous expression $\Omega_{\mathrm{m}} = \Omega_{\mathrm{de}}r(z)$. By evaluating at present time ($z=0$) the coincidence parameter we obtain $r(0) = [\Omega_{\mathrm{m,0}}(1+\omega_{0})]/[1-\Omega_{\mathrm{m,0}}(1+\omega_{0})]$, which differs from the standard result $\Omega_{\mathrm{m,0}}/[1-\Omega_{\mathrm{m,0}}]$, due to the interaction scheme. Using the equations (\ref{eq:q12}) and (\ref{eq:q22}) together with the relation between both densities, $\Omega_{\mathrm{m}} = \Omega_{\mathrm{de}}r(z)$, we can obtain the $Q$-term for the interaction 
\begin{equation}
    \frac{Q(z)}{3H^{3}_{0}E(z)} = -\frac{3\Omega_{\mathrm{de}}(z)r(z)}{1+r(z)}\left[\omega_{\mathrm{de}}(z)+\frac{1+z}{3}\frac{r'(z)}{r(z)}\right] = -\frac{3\Omega_{\mathrm{m}}(z)}{1+r(z)}\left[\omega_{\mathrm{de}}(z)+\frac{1+z}{3}\frac{r'(z)}{r(z)}\right],
    \label{eq:qterm4}
\end{equation}
where Eq. (\ref{eq:deint2}) must be considered in order to have an explicit expression for the interaction term. In order to show the behavior of the $Q$-term, we will consider the best fit value obtained by the Planck collaboration commented in the previous section. In Fig. (\ref{fig:planck}) we depict the interaction $Q$-term given in (\ref{eq:qterm4}). It is worthy to mention that the interaction term exhibit changes of sign at the past, becoming negative near the present time; these changes in the sign of the interaction term reveal some thermodynamics properties of the model, we will comment these features below.
\\

\begin{figure}[htbp!]
\centering
\includegraphics[width=8cm,height=6.5cm]{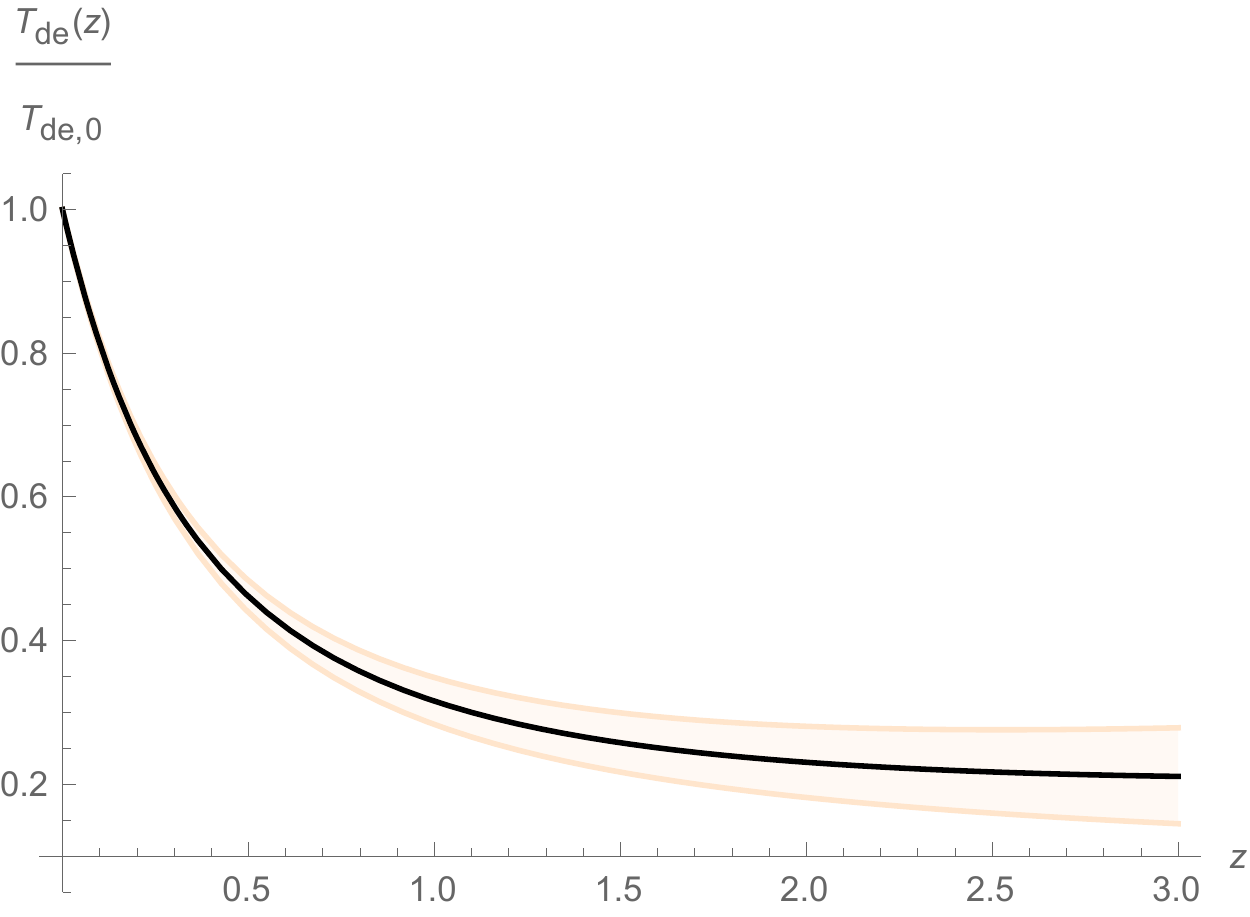}
\includegraphics[width=8cm,height=6.5cm]{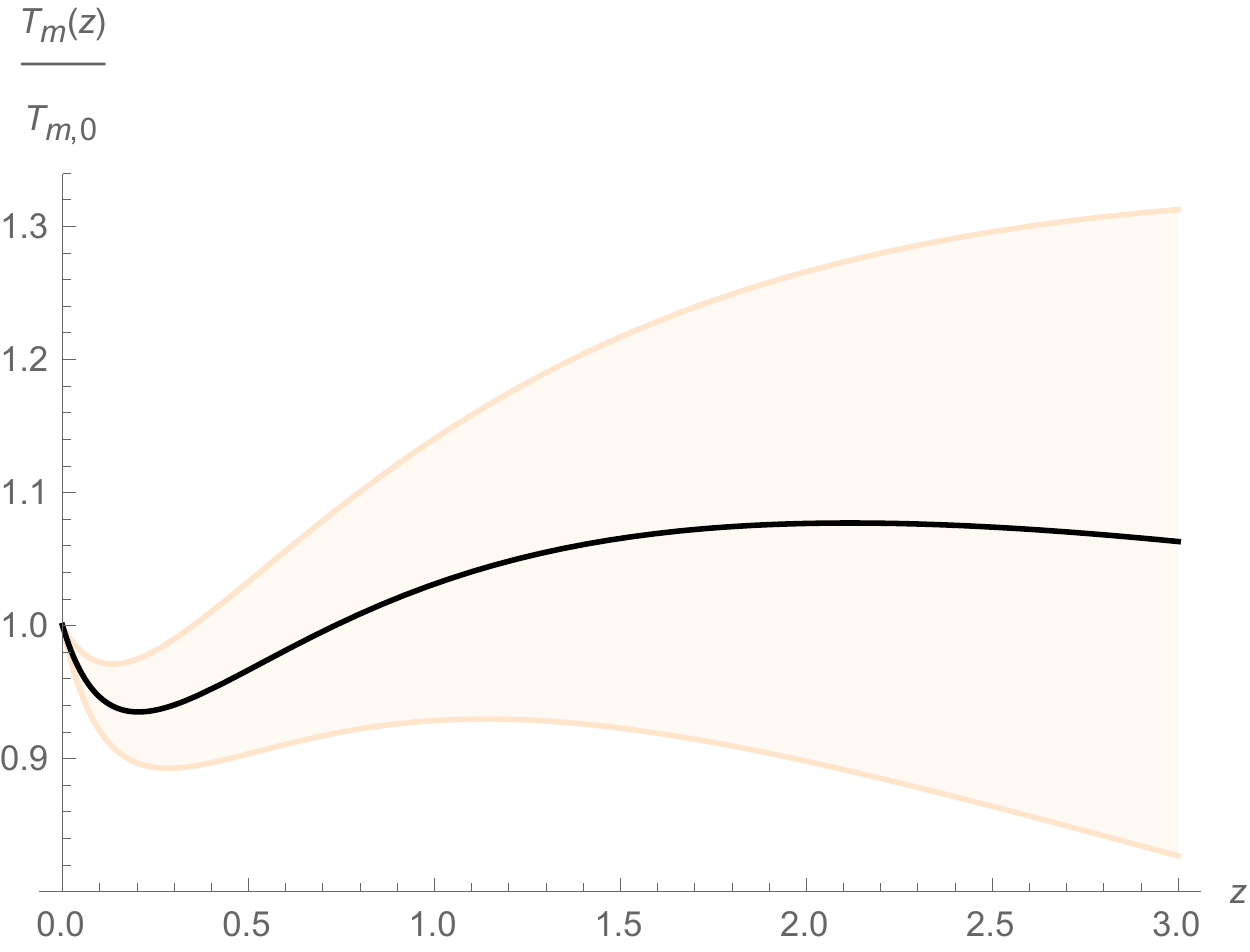}
\caption{Temperatures of the dark sector. The black solid line corresponds to the best fit values $\omega_{0}$ and $\omega_{a}$ in both panels, the shaded regions correspond to the errors propagation.}
\label{fig:temperatures}
\end{figure}

The form obtained for the above expressions are useful if we implement the effective temperatures method discussed in Ref. \cite{temperature}. From the continuity equations (\ref{eq:q12}) and (\ref{eq:q22}), we can write conveniently
\begin{align}
& \rho'_{\mathrm{de}} - 3 \left(\frac{1+\omega_{\mathrm{de,eff}}}{1+z} \right)\rho_{\mathrm{de}} = 0, \label{eq:q13}\\
& \rho'_{\mathrm{m}} - 3\left(\frac{1+\omega_{\mathrm{m,eff}}}{1+z} \right)\rho_{\mathrm{m}} = 0, \label{eq:q23}
\end{align} 
where we have defined
\begin{align}
& \omega_{\mathrm{de,eff}}(z) = \omega_{\mathrm{de}}(z) + \frac{Q(z)}{3\rho_{\mathrm{de}}H_{0}E(z)}, \label{eq:eff2}\\
& \omega_{\mathrm{m,eff}}(z) = -\frac{Q(z)}{3\rho_{\mathrm{m}}H_{0}E(z)}. \label{eq:eff1}
\end{align} 
In a single fluid description the evolution equation for the temperature is given by \cite{maartens}
\begin{equation}
    \frac{\dot{T}}{T} = -3H\frac{\partial p}{\partial \rho} \ \   \longrightarrow \ \  \frac{1}{T}\frac{dT}{dz} = 3\omega_{\mathrm{i,eff}}(1+z)^{-1},
\end{equation}
where the dot in the first equation denotes derivative with respect to time. On the other hand, we will consider that the pressure and energy density of each component will be related through the effective parameter state by means of a barotropic equation of state. Therefore, solving the previous expression for the temperature one gets
\begin{equation}
    T_{\mathrm{i}}(z) = T_{\mathrm{i,0}}\exp\left\lbrace 3\int^{z}_{0} \omega_{\mathrm{i,eff}}(x) d\ln (1+x)\right\rbrace,
\end{equation}
being $\mathrm{i} = \mathrm{de}, \mathrm{m}$. Taking into account the equations (\ref{eq:deint2}), (\ref{eq:qterm4}), (\ref{eq:eff2}) and (\ref{eq:eff1}) in the above equation, we can obtain for each temperature the following expressions
\begin{equation}
    T_{\mathrm{de}}(z) = T_{\mathrm{de,0}}(1+z)^{-\alpha}\left\lbrace 1 + \frac{[\Omega_{\mathrm{m,0}}(1+\omega_{0}+\omega_{a})-1]}{\Omega_{\mathrm{m,0}}(1+\omega_{0})-1}z \right\rbrace^{-\beta}\exp\left(3\omega_{a}\frac{z}{1+z}\right),
    {\label{eq:ttde}}
\end{equation}
for simplicity in the notation we have defined the constants $\alpha$ and $\beta$, their explicit expressions are given as follows
\begin{equation*}
    \alpha := \frac{6-\Omega_{\mathrm{m,0}}+3\Omega_{\mathrm{m,0}}(\omega_{0}+\omega_{a})}{\Omega_{\mathrm{m,0}}}, \ \ \ \beta := \frac{6-7\Omega_{\mathrm{m,0}}+\Omega^{2}_{\mathrm{m,0}}(1+\omega_{0}+\omega_{a})}{\Omega^{2}_{\mathrm{m,0}}(1+\omega_{0}+\omega_{a})-\Omega_{\mathrm{m,0}}}.
\end{equation*}
Finally, for the dark matter component we have
\begin{equation}
    T_{\mathrm{m}}(z) = T_{\mathrm{m,0}}(1+z)^{-3(\omega_{0}+\omega_{a})}\left[\frac{1+\omega_{0}}{1+\omega_{0}+\omega_{a}\frac{z}{1+z}}\right]\exp\left(3\omega_{a}\frac{z}{1+z}\right).
\label{eq:ttdm}
\end{equation}
In Fig. (\ref{fig:temperatures}) we show the quotients, $T_{\mathrm{i}}(z)/T_{\mathrm{i,0}}(z)$, obtained from the temperatures (\ref{eq:ttde}) and (\ref{eq:ttdm}), using the best fit values obtained from the statistical analysis. As can be seen on the left panel, for dark energy the temperature will reach its maximum value at present time, $z=0$. However, for the dark matter component, its temperature exhibits (not shown in the plot) a pathological behavior. Around some future value for the redshift, $z\approx -0.245$, the dark matter temperature increases abruptly and rapidly decreases to negative values, this behavior is due to the term in square brackets of Eq. (\ref{eq:ttdm}). Therefore, this interacting scenario could be applicable only to describe the Universe from past until present time. If we only consider the aforementioned period for the cosmic expansion, the behavior of both temperatures is distinct, while the dark energy sector has a temperature that increases along the cosmic evolution, the dark matter temperature around $z\approx 2$ changes its increasing tendency and around $z=0.2$ its value increases again.    
\\

\begin{figure}[htbp!]
\centering
\includegraphics[width=13cm,height=6.5cm]{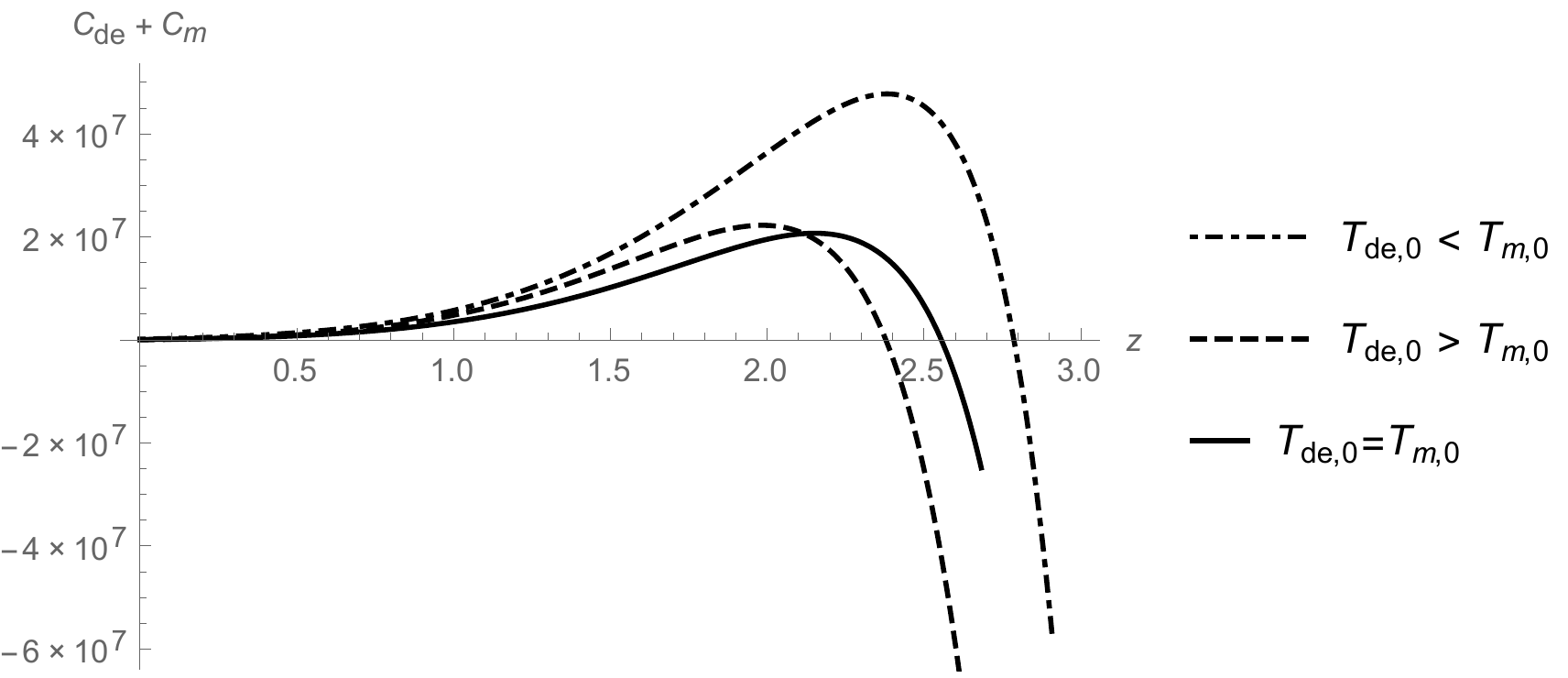}
\caption{Sum of heat capacities using the best fit values for the parameters of the model.}
\label{fig:capacities}
\end{figure}

As discussed in Ref. \cite{pena}, the change in the sign of the interaction term reveal the existence of possible phase transitions since the sign of the heat capacities of the components also change, this also implies changes in the temperatures of the components. For $Q>0$ the heat capacities of both components can be written as \cite{pena}
\begin{eqnarray}
    C_{\mathrm{de}} &=& -\frac{Q}{\Delta T_{\mathrm{de}}} < 0, \ \ \ \mbox{if} \ \ \ \Delta T_{\mathrm{de}} > 0,\label{eq:heat1}\\
    C_{\mathrm{m}} &=& \frac{Q}{\Delta T_{\mathrm{m}}} > 0, \ \ \ \mbox{if} \ \ \ \Delta T_{\mathrm{m}} > 0. \label{eq:heat2}
\end{eqnarray}
For instance in the interval $2.1 < z < \infty$, we have $Q<0$ and $\Delta T > 0$ for both components, therefore $C_{\mathrm{de}} > 0$ and $C_{\mathrm{dm}} < 0$. On the other hand, for the interval $0.2 < z < 2.1$ we have $Q>0$ and $\Delta T_{\mathrm{m}} < 0$, $\Delta T_{\mathrm{de}} >0$, in this case the heat capacities satisfy the conditions $C_{\mathrm{de}} < 0$ and $C_{\mathrm{dm}} < 0$. Finally, for the interval $0 < z < 0.2$ we have $Q<0$ and we again obtain $C_{\mathrm{de}} > 0$ and $C_{\mathrm{dm}} < 0$. As can be seen, for this interacting scheme the dark matter sector temperature is more sensitive to the changes of sign of the interaction term. On the other hand, we do not visualize phase transitions at the past and present time; given the description above for the behavior of the dark matter temperature at the future, we could expect a future phase transition, but as also was commented, this stage is not considered in our description. In Fig. (\ref{fig:capacities}), we plot the sum of the heat capacities (\ref{eq:heat1}) and (\ref{eq:heat2}), their definitions imply the consideration of Eqs. (\ref{eq:qterm4}), (\ref{eq:ttde}) and (\ref{eq:ttdm}). Note that we depend on the initial value of both temperatures, in the graphic we show three possible cases $T_{\mathrm{de,0}} = T_{\mathrm{m,0}}$, $T_{\mathrm{de,0}} < T_{\mathrm{m,0}}$ and $T_{\mathrm{de,0}} > T_{\mathrm{m,0}}$. According to \cite{pena2}, the equilibrium condition is reached only if the sum of heat capacities is negative along the cosmic evolution. For the three cases considered here it is found that the system was taken out from equilibrium, i.e., at certain value for the redshift at the past the sum of heat capacities becomes positive and keeps that way until present time.    
\\

\begin{figure}[htbp!]
\centering
\includegraphics[width=13cm,height=6.5cm]{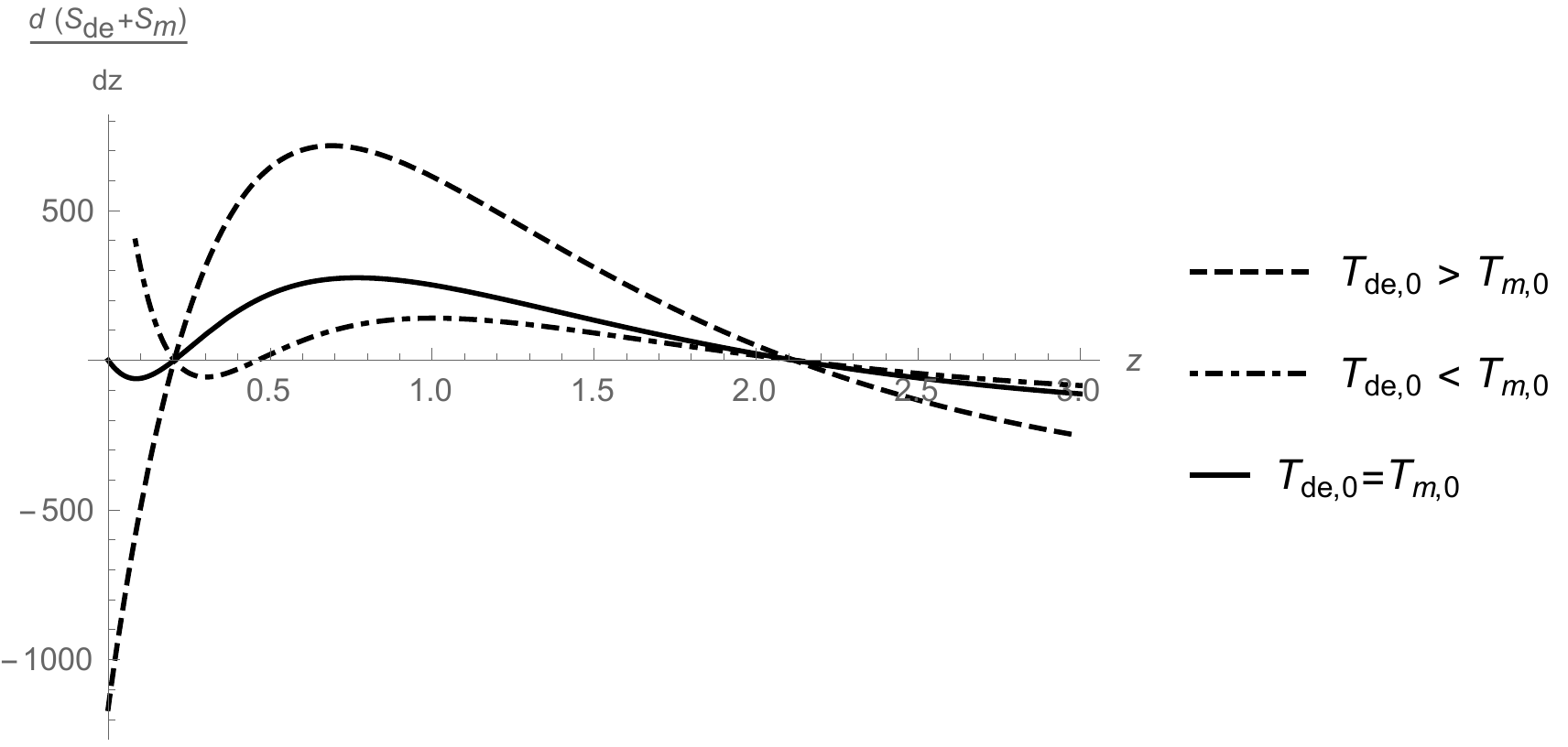}
\caption{Behavior for the total entropy of the system using the best fit values for the parameters of the model.}
\label{fig:secondlaw}
\end{figure}

From the second law we have
\begin{equation}
    TdS = d(\rho V) + pdV,
    \label{eq:second}
\end{equation}
being $V$ the Hubble volume defined as $V(z)=V_{0}(a/a_{0})^{3}=V_{0}(1+z)^{-3}$, in this case the quantity $\rho V$ denotes the internal energy. If we consider the Eqs. (\ref{eq:q12}), (\ref{eq:q22}) in the above equation, we can write for each component of the dark sector
\begin{equation}
    -\frac{T_{\mathrm{m}}}{V}\frac{dS_{\mathrm{m}}}{dz} = \frac{Q}{H_{0}E(z)(1+z)} = \frac{T_{\mathrm{de}}}{V}\frac{dS_{\mathrm{de}}}{dz},
    \label{eq:total}
\end{equation}
then the entropy associated to each component is not constant, the adiabatic condition ($S = \mbox{constant}$) for the cosmic evolution is recovered for null interaction, $Q=0$. From Eq. (\ref{eq:total}) one gets the following condition
\begin{equation}
    T_{\mathrm{m}}dS_{\mathrm{m}} + T_{\mathrm{de}}dS_{\mathrm{de}} = 0,
\end{equation}
which leads to
\begin{equation}
    \frac{d}{dz}(S_{\mathrm{m}}+S_{\mathrm{de}}) = -\left(-1+\frac{T_{\mathrm{de}}}{T_{\mathrm{m}}} \right)\frac{dS_{\mathrm{de}}}{dz}.
    \label{eq:entropy}
\end{equation}
Thus, from Eq. (\ref{eq:total}) we can observe that the behavior of the $Q$-term dictates the sign of $dS_{\mathrm{de}}/dz$, in consequence the fulfillment of the second law, $dS/dz < 0$, by this model will depend on the behavior of the following three physical quantities: $T_{\mathrm{de}}$, $T_{\mathrm{m}}$ and the interaction $Q$-term. In Fig. (\ref{fig:secondlaw}) the results for the derivative of the total entropy given in (\ref{eq:entropy}) can be found for three different initial conditions for the components temperatures. As can seen, the second law it is not guaranteed during all the cosmic expansion and this is consequence of the dependence of Eq. (\ref{eq:entropy}) on the interaction term, as can be seen in (\ref{eq:total}); which exhibits changes of sign. Only the possibilities $T_{\mathrm{de,0}} > T_{\mathrm{m,0}}$ and $T_{\mathrm{de,0}} = T_{\mathrm{m,0}}$, guarantee the fulfillment of the second law at present time.

\section{Concluding remarks}
\label{sec:final}
In this work we explored two possible cosmological scenarios for the reconstructed approach of mimetic gravity. As discussed previously, the role of the Lagrange multiplier becomes relevant since enforces an important constraint for the mimetic field and besides its interpretation as emergent matter sector avoids the introduction of extra fields in the description; a first interesting feature of this model is the deviation from the standard behavior that generates the potential on the Lagrange multiplier, then such deviation reveals a rich structure in the model; such kind of deviations for the matter sector are generally present in interacting models, this latter possibility was explored here. Secondly, in our description the Lagrange multiplier together with the potential were reconstructed, then we do not place them by hand. The first scenario corresponds to the inclusion of matter production effects in the mimetic description, as far as we know this is the first time that such effects are considered in the mimetic approach, these effects are expected to contribute on the cosmic expansion by means of an effective pressure; the second case corresponds to an interaction scheme for the dark sector that emerges from mimetic gravity, therefore the mimetic scheme can be seen as an unified model for dark matter and dark energy. The free parameters of the model were constrained for each aforementioned case with the use of recent cosmological data. It is worthy to mention that at $1\sigma$ the lower bound of the parameter $\omega_{0}$\footnote{We must have in mind that this parameter represents the present time value for the parameter state in the CPL parametrization.} for the matter production case can cross to the phantom zone. This was confirmed at effective level, the parameter state exhibits a transitory behavior, i.e., the cosmic expansion can be driven by a quintessence dark energy fluid for a while and eventually a phantom scenario dominates. It is found that a future singularity for the normalized Hubble parameter can take place only at the far future, $z=-1$, this is known as little rip singularity. In summary, the matter production effects for this approach could describe the nature of the observable Universe. For the interaction scenario the constrained value for $\omega_{0}$ will represent a quintessence dark energy. A relevant feature of the mimetic approach is the value for $h$ constrained in each case studied here, these values are closer to the Planck results than those reported by SH0ES, then, the approach considered in this work could provide a viable alternative to alleviate the $H_{0}$ tension.\\

On the other hand, a novelty of the reconstructed mimetic gravity is the emerging interacting scenario for the dark sector. In this work we focused on the thermodynamics characteristics of this interacting description. As commented in the work, the region of validity for this description, which results to be valid only from past to present time, was established by the behavior of the temperature associated to the dark matter component in order to avoid some pathological situations from the thermodynamics point of view, this delimitation also guarantees that the resulting scenario avoids the future singularity inherent to the CPL parametrization. In our description we do not assume an Ansatz for the interaction $Q$-term, the model itself provides a construction for this term. Using the constrained values for the parameters of the model, we observe changes in the sign of $Q$ along the cosmic evolution, in general these changes in its sign can induce phase transitions or changes in the temperatures of the components. Although there are some changes in the heat capacities of the components, we do not observe phase transitions for this model but we can note that the dark matter temperature is sensitive to the changes of $Q$. This thermodynamics description depends on the initial values of the temperatures but as far as we know, there are not definite results coming from observations that could reveal some specific values for these initial temperatures yet, thus we can consider three cases: equal initial temperatures and one initial temperature greater than the other and vice versa. Under these three cases, we show that the sum of the heat capacities of the components of the dark sector is negative at past and at specific value of the redshift turns positive, this means that the thermal equilibrium was lost at a specific moment at the past and such condition keeps until present time, {\it today} the components of the dark sector are out of thermal equilibrium. Finally, as it is well known, the interaction scenario leads to non adiabatic expansion, which seems to be more consistent than the one obtained in the $\Lambda$CDM model. Therefore the second law can be explored in these kind of models, in this work we found that the second law is not guaranteed for the whole cosmic expansion but if we look only at present time, its fulfillment is obtained if we consider equal initial temperatures for the components of the dark sector or if we consider that the initial value for the temperature of dark energy is greater than the initial value temperature of dark matter. We consider that whether or not the second law is obeyed in the interaction description, is directly related to the behavior exhibited by $Q$ and in our case the interaction term has an alternating nature. Note that from the three possibilities that we have considered in this work for the initial values of the temperatures, only in two cases the regions where the second law is well defined coincide with the regions where the interaction term is negative.    

\section*{Acknowledgments}
M.C. work has been supported by S.N.I. (CONACyT-M\'exico).

\end{document}